\documentclass[11pt,twoside]{article}
\usepackage{asp2010}

\resetcounters

\begin{document}

\title{Simulations of barred galaxies in triaxial dark matter haloes: \\the effects of gas}
\author{Rubens~E.~G.~Machado$^{1,2}$, E.~Athanassoula$^1$, and S.~A.~Rodionov$^1$
\affil{$^1$ Aix Marseille Universit\'e, CNRS, LAM (Laboratoire d'Astrophysique de Marseille) UMR 7326, 13388, Marseille, France}
\affil{$^2$Instituto de Astronomia, Geof\'isica e Ci\^encias Atmosf\'ericas,\\ Universidade de S\~ao Paulo, R. do Mat\~ao 1226, 05508-090 S\~ao Paulo, Brazil}}

\begin{abstract}
The baryonic discs of galaxies are believed to alter the shapes of the dark matter haloes in which they reside. We perform a set of hydrodynamical N-body simulations of disc galaxies with triaxial dark matter haloes, using elliptical discs with a gaseous component as initial conditions. We explore models of different halo triaxiality and also of different initial gas fractions, which allows us to evaluate how each affects the formations of the bar. Due to star formation, models of all halo shapes and of all initial gas fractions reach approximately the same gas content at the end of the simulation. Nevertheless, we find that the presence of gas in the early phases has important effects on the subsequent evolution. Bars are generally weaker for larger initial gas content and for larger halo triaxiality. The presence of gas, however, is a more efficient factor in inhibiting the formation of a strong bar than halo triaxiality is.
\end{abstract}

\section{Introduction}

Cosmological simulations generally show that dark matter haloes are not spherical, but rather triaxial (or prolate). The baryonic discs of galaxies are expected to affect the shapes of their haloes. Here we aim to explore how the presence of a gaseous component in the disc influences bar formation within elliptical discs embedded in triaxial haloes. These simulations are similar to those of \citet{MachadoAthanassoula2010}, but now a fraction of the disc mass is initially in the form of gas. We present here one spherical halo and one triaxial halo, with axes initially in the following ratios: 1:1:1 (halo~1); and 1:0.6:0.4 (halo~3). In each case, we include discs having five different initial gas fractions: 0, 20, 50, 75 and 100\%. Initial conditions are created with the iterative method described in \citet{Rodionov2009} and \citet{Rodionov2011}. To calculate the evolution, we use a version of the {\sc gadget2} code \citep{Springel2005} that includes star formation \citep{SpringelHernquist2002,SpringelHernquist2003}.

\section{Simulation results}

\articlefigure[width=\textwidth]{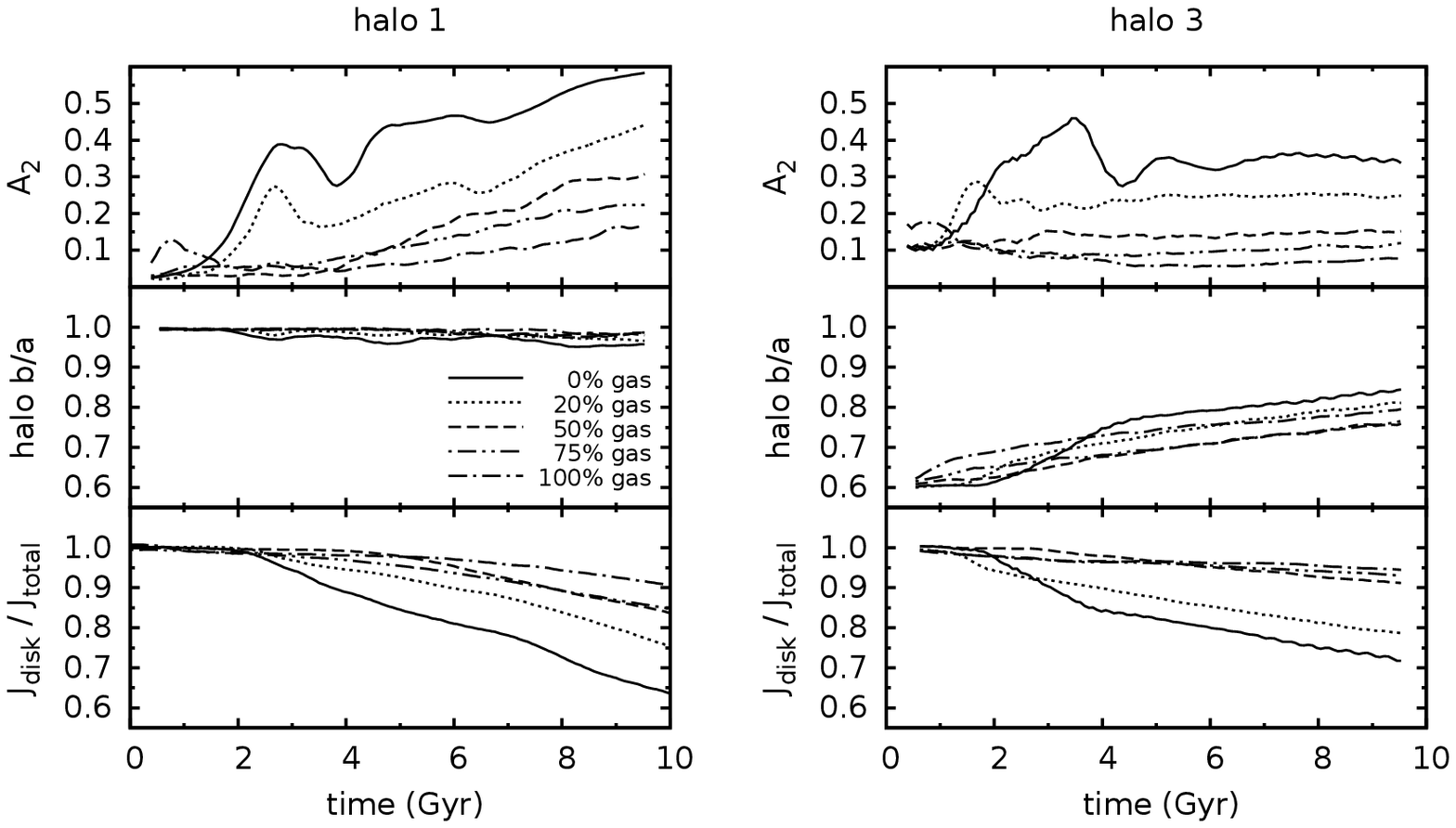}{Fig1}{Evolution of the bar strength (top row), halo equatorial shape (middle row) and fractional disc angular momentum (bottom row), for haloes of two different shapes. Line types indicate each model's initial gas fraction.} 

Models with high initial gas fraction develop bars that are substantially weaker (top row of Fig.~\ref{Fig1}) and in such weakly barred models, the transfer of angular momentum from the disc to the halo \citep{Athanassoula2003} is correspondingly smaller (bottom row of Fig.~\ref{Fig1}). We find the strongest bar in the model where the halo is spherical and in which there is no gas. Bars are generally weaker for larger initial gas content and for larger halo triaxiality. The presence of gas, however, is a more efficient factor in inhibiting the formation of a strong bar than halo triaxiality is. Bar formation causes the haloes to become more axisymmetric (middle row of Fig.~\ref{Fig1}). We also observe the formation of a bar in the halo component \citep[as in][]{Athanassoula2007} in cases of strong stellar bars, but if the bar is weak the central part of the halo becomes circularised. Rotation of the disc-like halo particles \citep{Athanassoula2007} is also present in these simulations with gas, and we find a clear correlation between bar strength and the peak tangential velocity of these rotating halo particles, in the sense that the rotation is more important in strongly barred models. 

\acknowledgements RM acknowledges support from Brazilian agencies FAPESP (05/04005-0) and CAPES (3981/07-0), and from the French Ministry of Foreign and European Affairs (bourse Eiffel). This work was partly supported by grant ANR-06-BLAN-0172.

\bibliography{Machado_R}

\end{document}